\documentclass[twocolumn,showpacs,prl,superscriptaddress]{revtex4}
\usepackage{amssymb}
\usepackage{amsfonts}
\usepackage{amsmath}
\usepackage{graphicx}
\usepackage{graphics}
\usepackage{bm}
\usepackage{mathdots}
\usepackage{calc}
\usepackage{dcolumn}

\newcommand{\tr}{{\mbox{Tr}}}
\usepackage{pstricks}
\usepackage{pst-plot}
\begin{document}
\title{Quantum contextuality  and joint measurement of three observables of a qubit 
}

\author{Sixia Yu}
\affiliation{Centre for Quantum Technologies, National University of Singapore, 2 Science Drive 3, Singapore 117542}
\author{C.H. Oh}
\affiliation{Centre for Quantum Technologies, National University of Singapore, 2 Science Drive 3, Singapore 117542}
\affiliation{Physics Department, National University of Singapore, 2 Science Drive 3, Singapore 117542}

\begin{abstract}
Whereas complementarity manifests itself via two incompatible observables, quantum contextuality can only be revealed via the joint measurements among at least three observables. By incorporating unsharp measurements and joint measurements into a realistic model, we reestablish an inequality due to Liang, Spekkens, and Wiseman rigorously based  on the assumption of noncontextuality alone. Its violation therefore unambiguously pinpoints the quantum contextuality of a two-level system. The maximal violation is attained by  three triplewise jointly measurable observables that are pairwise jointly measured in an incompatible way. We also present the necessary and sufficient condition of triplewise joint measurability of three unbiased observables of a qubit.  \end{abstract}
\maketitle

Complementarity and quantum contextuality are two fundamental essential elements of quantum theory, both originate in the fact that there are incompatible observables that cannot be measured jointly in a single measurement apparatus. As stated by Bohr's complementarity principle  \cite{bohr}, there are mutually exclusive but equally real aspects of quantum systems. Two exclusive properties or two incompatible observables are enough to demonstrate complementarity, e.g., wave versus particle and momentum versus position. Quantitatively, the complementarity can be demonstrated by various kinds of uncertainty relationships for preparation \cite{heis, rs,yu0} as well as for the joint measurement of two noncommuting observables \cite{werner, busch13, ozawa,englert}.

Unlike complementarity, quantum contextuality can only be revealed by the joint measurements among at least three observables, e.g., via the pairwise joint measurements.   A joint measurement of two or more observables defines a measurement context and at least two different contexts must be present to demonstrate the contextuality. By non-contextuality we mean that the outcomes of a faithful measurement  of a given observable are predetermined regardless of what other compatible observables might be measured along.  For systems with three or more distinguishable states,  Kochen and Specken (KS) \cite{ks}, as well as by Bell \cite{bell} proved quantum contextuality by showing that non-contextual realistic models cannot reproduce all the predictions of quantum mechanics via mere logical contradictions. With the help of non-contextuality inequalities \cite{cab3,bent}, or KS inequalities, quantum contextuality can be put to experimental tests, just like Bell inequalities. There are also state-independent proofs  \cite{yu,yu1} that do not arise from KS-type logical contradictions.

In the case of two-level systems, or qubits, proofs of quantum contextuality  inevitably involve some additional assumptions other than non-contextuality. For examples, Cabello and Nakamura \cite{cab2} tried to extract a logical contradiction of KS type by assuming the outcome determinism for unsharp measurements. However, a non-contextual model \cite{allen,ks21} was found to explain the contradictions away. Accardi \cite{accardi} and Fujikawa \cite{fuji} used the conditional probabilities of sequential measurements, assuming the existence of their counterparts in quantum theory,  in their arguments.  Busch \cite{busch}, as well as Caves etal, \cite{caves} proved Gleason's theorem for qubit by assuming the additivity of unsharp measurements. 

Notably, Spekkens \cite{spekkens} formulated a kind of generalized notion of measurement contextuality, assuming certain linearity for unsharp measurement. Later on, based on this notion, Liang, Spekkens, and Wiseman (LSW) \cite{lsw} derived an inequality  on the average anti-correlations  in the Specker's scenario \cite{specker} in which three pairwise joint measurements of three observables  define three measurement contexts. Recently Runjawl and Ghosh \cite{kg}  found out a violation to LSW's inequality in a qubit. However, because of some additional assumptions used in the original derivation of LSW's inequality, the violation to LSW's inequality does not clearly pinpoint the quantum contextuality of a qubit.

The purpose of this Letter is twofold. One is classical: we shall at first model unsharp measurements and joint measurements in a non-contextual realistic model and then re-derive LSW's inequality based rigorously on the assumption of non-contextuality alone.  One is quantum: we shall derive the maximal violation to LSW's inequality by a qubit and the necessary and sufficient condition for three unbiased observables of a qubit to be triplewise joint measurable. Contrary to the customary expectations, it is the compatibility of pairwise joint measurements instead of triplewise joint measurability that is  relevant to the quantum contextuality of a qubit.

Quantum theory can be regarded a special kind of operational probabilistic theory, making statistical predictions on possible measurements. The issue of contextuality arises when one tries to attribute properties to the system independent of the measurements, i.e., to understand the statistical predictions from a  non-contextual and realistic point of view. In a non-contextual realistic model all observables have realistic  values predetermined by some hidden variables, denoted collectively as $\lambda$, distributed according to some probability distribution $\varrho_\lambda$ normalized to 1. Measurements are physical processes capable of revealing these predetermined values. Some measurements are faithful, called here as {\it sharp} measurements, and some might not be so faithful, called here as {\it unsharp} measurements. How a measuring apparatus responses to the predetermined values is recorded by the so-called {\it response function}, which was first introduced by Spekkens \cite{spekkens}. A sharp measurement of an observable $A$ yields outcomes that are identical to the predetermined values $A(\lambda)$ and therefore it has an ideal {response function}  $\chi_\mu[A(\lambda)]$, which equals to 1 if $A(\lambda)=\mu$ and 0 otherwise.  The probability of obtaining outcome $k$ by a sharp measurement of $A$ reads
$P(\mu|A)=\int d\lambda \varrho_\lambda \chi_\mu[A]:=\langle \chi_\mu[A]\rangle.$ Since the predetermined values of an observable are assumed to be non-contextual, the response functions of sharp measurements are also non-contextual.

In an unsharp measurement, however, the measuring apparatus might give wrong responses to the actual predetermined values, e.g., there might be a nonzero probability $P_{\mu|\mu^\prime}(\lambda)$ of obtaining outcome $\mu$ when the observable $A$ is predetermined to have value $\mu^\prime$. Obviously it  holds $\sum_\mu P_{\mu|\mu^\prime}(\lambda)=1$ for any given $\mu^\prime$ and $\lambda$. In this case the probability of obtaining outcome $\mu$ reads
\begin{equation}\label{rf3}
P(\mu|\tilde A)=\sum_{\mu^\prime}\int d\lambda\ \varrho_\lambda P_{\mu|\mu^\prime}(\lambda)\chi_{\mu^\prime}[A]:=
\langle \tilde\chi_{\mu}[A]\rangle
\end{equation}
where $\tilde\chi_\mu[A]=\sum_{\mu^\prime} P_{\mu|\mu^\prime}(\lambda)\chi_{\mu^\prime}[A]$ is defined to be the response function of an unsharp measurement of $A$. In general, the response function $\tilde\chi_\mu[A]$ of a measurement of $A$ with outcome $\mu$ is defined to be a function  of the hidden variable $\lambda$ satisfying 
 \begin{equation}
\mbox{RF1. }\tilde\chi_\mu[A]\ge0,\quad \mbox{RF2. }\sum_\mu\tilde\chi_\mu[A]=1
\end{equation}
such that RF3. (c.f. Eq.(\ref{rf3})) {\it the probability of obtaining outcome $\mu$ is given by the average of the response function}. Condition RF3 defines the response functions whereas conditions RF1 and RF2 are justified by the fact that the probability is nonnegative and normalized, respectively, for any distribution $\varrho_\lambda$ of hidden variables. 
For unsharp measurements the response functions may differ from the predetermined non-contextual values and can even be contextual in a non-contextual model as will be shown below. In comparison, Spekkens \cite{spekkens} assumed that the response functions are  predetermined and  non-contextual albeit non-deterministic values of observables. 

By a {\it joint measurement} of two observables $A_1$ and $A_2$ we mean any unsharp measurement that outputs a joint probability distribution of measurement results of $A_1$ and $A_2$ in any distribution $\varrho_\lambda$ of the hidden variables. The response function $\tilde\chi_{\mu\nu}[A_{12}]$ of the joint measurement has two response functions of $A_1$ and $A_2$ as marginals, i.e., 
\begin{equation}
\tilde\chi_{\mu}[A_1]=\sum_\nu\tilde\chi_{\mu\nu}[A_{12}],\quad \tilde\chi_{\nu}[A_2]=\sum_\mu\tilde\chi_{\mu\nu}[A_{12}].
\end{equation}
 The joint measurement of three or more observables can be defined similarly.
Unlike quantum cases, in a non-contextual realistic model all observables are jointly measurable since the product of all the response functions defines a joint measurement. 

A joint measurement defines a measurement context and different measurement contexts may be incompatible even classically.
Three pairwise joint measurements  $\tilde\chi_{\mu\nu}[A_{jk}]$  of three observables $A_k$ with $j< k$ and $j,k=1,2,3$ are   {\it compatible} if there exists a joint measurement $\tilde\chi_{\mu\nu\tau}[A_{123}]$ that has those three response functions of pairwise joint measurements as marginals, e.g., 
\begin{equation}
\tilde\chi_{\mu\nu}[A_{12}]=\sum_\tau\tilde\chi_{\mu\nu\tau}[A_{123}].
\end{equation}

Specifically, we consider in what follows binary observables taking values $\pm1$. The response function of a sharp measurement of a binary observable is given by $\chi_\mu[A]=({1+\mu A})/2$. 
In an unsharp measurement of $A$ there might be a probability $P_+(\lambda)$ or $P_-(\lambda)=1-P_+(\lambda)$ of obtaining an outcome $\mu=\pm1$ if the predetermined value of $A$ is actually $\mu$ or $\bar\mu=-\mu$, respectively.  Denote by $\eta(\lambda)=P_+(\lambda)-P_-(\lambda)$ the {\it local unsharpness} and the corresponding response function reads
\begin{equation}
\tilde\chi_\mu[A]=P_+(\lambda)\chi_\mu[A]+P_-(\lambda)\chi_{\bar\mu}[A]=\frac{1+\mu\eta(\lambda) A}2.
\end{equation}
The {\it (global) sharpness} $\eta$ of an unsharp measurement can be understood in a theory independent fashion as 
\begin{equation}
\eta=\min\max_{A=\pm}\left|P(+|\tilde A)-P(-|\tilde A)\right|
\end{equation}
with minimization taken over all possible states in which observable $A$ has definite values. In a realistic model $\eta=\min\langle |\eta(\lambda)|\rangle$ over all distributions $\varrho_\lambda$ of $\lambda$. 

Given three binary observables $A_1,A_2$ and $A_3$, the most general pairwise joint measurement of observable $A_j$ and $A_k$ has a response function
\begin{equation}\label{rf2}
\tilde\chi_{\mu\nu}[A_{jk}]=\frac{1+\mu\eta_j(\lambda)A_j+\nu\eta_k(\lambda)A_k+\mu\nu C_{jk}}4
\end{equation}
where $C_{jk}$ is an observable whose predetermined values satisfy
$
1\pm C_{jk}\ge| \eta_j(\lambda) A_j\pm\eta_k(\lambda)A_k|
$ to ensure  $\tilde\chi_{\mu\nu}[A_{jk}]\ge0$. The anti-correlation, i.e.,  the probability of obtaining different outcomes in a joint measurement, is given by the average of $(1+C_{jk})/2$.

{\it Theorem 1. Three pairwise joint measurements of three binary observables are compatible if and only if
\begin{equation}\label{t1}
1-|C_{13}-C_{23}|\ge C_{12}\ge|C_{13}+C_{23}|-1.
\end{equation}}%

Proof is given in Supplemental Material \cite{sm} and we note an interesting similarity to Accardi's inequality on conditional probabilities \cite{accardi}. For three compatible pairwise joint measurements,  because of the triplewise joint measurement, all the outcomes of pairwise joint measurements can be accounted for in a non-contextual manner, i.e., they are
 determined by the hidden variables alone and independent of which observables might be measured along. In other words, the probability of giving a false response to an actual value is independent of what other observables that might be measured along.   However  three observables can also be pairwise jointly measured in an incompatible manner. For an example, the pairwise joint measurements of three observables given by $\eta_k=\eta<1/2$ for $k=1,2,3$ and $C_{jk}=\eta(1+A_jA_k)-1$ violate the  condition Eq.(\ref{t1}) and therefore are incompatible. In this case a non-contextual account for all the long-run statistics is impossible because of the absence of a joint probability distribution. Thus even  a non-contextual realistic model may exhibit measurement contextuality. 

However this measurement contextuality induced by unsharp measurements cannot account for the quantum contextuality. 
To show this we consider  the {\it average anti-correlation}, i.e., the average probability of obtaining different outcomes,  in three pairwise joint measurements
\begin{eqnarray}\label{r3}
R_3=\frac13\sum_{j<k}\sum_{\mu=\pm}P(\mu,-\mu|A_{jk})
\end{eqnarray}
which is first introduced by LSW \cite{lsw}. 
For compatible pairwise joint measurements,  using Theorem 1, it holds $R_3\le 2/3$  \cite{lsw}. A violation $R_3>2/3$ does not mean that those three observables are not triplewise jointly measurable. Instead, it means that these three observables are pairwise-jointly measured in an incompatible way.

{\it Theorem 2. In a non-contexutal realistic model the average anti-correlation of three unsharp measurements with sharpness $\eta_1\ge\eta_2\ge\eta_3$ satisfies}
\begin{equation}\label{t2}
R_3\le 1-\frac{\eta_1}3.
\end{equation}

Proof is given in Supplemental Material \cite{sm}. In appearance this is just a trivial generalization of LSW's inequality to the case of unequal sharpness. However there are two main differences in their derivations. First, in its original proof \cite{lsw}  the response function of unsharp measurement is obtained on the assumption of certain linearity of response functions,  which may not hold in some non-contextual models \cite{hermans}, so that only global sharpness is considered.  Second, in its original proof \cite{lsw} only a subset of response functions Eq.(\ref{rf2}), in which $C_{jk}$ is implicitly assumed to be a function of $A_j$ and $A_k$, was taken into account. Here we have taken into account all possible response functions conforming to the long-run statistics and, because of the local sharpness, the response function needs not to be linear. As a result we have established LSW's inequality rigorously based on the assumption of non-contextuality alone, i.e., it is valid for any non-contextual model admitting unsharp measurements and joint measurements. All relevant quantities such as sharpness, joint measurements, anti-correlations  are also well defined in quantum theory. 
 
Quantum mechanically the most general measurement is a positive operator valued measure (POVM), a set of positive operators $\{O_\mu\ge0\}$ summed up to the identity, playing the role of response function.   Two observables $O^1_\mu$ and $O^2_\nu$, are {\it jointly measurable} if there exists an observable $\{M_{\mu\nu}^{12}\}$ having two given POVMs as marginals, i.e., $O^1_\mu=\sum_{\nu}M_{\mu\nu}^{12}$ and $O^2_\nu=\sum_{\mu}M_{\mu\nu}^{12}$. 
 Three observables $\{O_{\mu}^k\}$ with $k=1,2,3$ are called {\it triplewise jointly measurable} if there is a  joint observable $\{M_{\mu\nu\tau}\}$ having the three given observables as marginals, e.g.,
$O_\mu^1=\sum_{\nu\tau} M_{\mu\nu\tau}$.  Three pairwise joint measurements $\{M^{ij}_{\mu\nu}\}_{i<j}$ are {\it compatible} if there exists a triplewise joint observable $\{M_{\mu\nu\tau}\}$ such that these three pairwise measurements arise as marginals, e.g., $
M_{\mu\nu}^{12}=\sum_{\tau}M_{\mu\nu\tau}$. Obviously triplewise jointly measurable observables are  pairwise jointly measurable and three observables having compatible pairwise joint measurements are triplewise jointly measurable. However, triplewise joint measurable observables may have incompatible pairwise joint measurements. 

An unbiased observable of a qubit refers to a two-outcome POVM 
$\{O_\pm(\vec\lambda)=\frac12({1\pm\vec \lambda\cdot \vec \sigma})\}$
with $\eta=|\vec\lambda|\le1$ being exactly the global sharpness. It is unbiased in the sense that the outcomes of the measurement are purely random if the
system is in the maximally mixed state. The necessary and sufficient condition for the joint measurability of two most general unsharp observables of a qubit has been found \cite{liu,jm,busch4,stano2,wolf}. Two unbiased observables $\{O_\pm(\vec\lambda_{i,j})\}$ are joint measurable if and only if \cite{busch2} $$H_{ij}:=1-|\vec\lambda_i|^2-|\vec\lambda_j|^2+(\vec\lambda_i\cdot\vec\lambda_j)^2\ge0.$$

\begin{figure}
\includegraphics[scale=0.65]{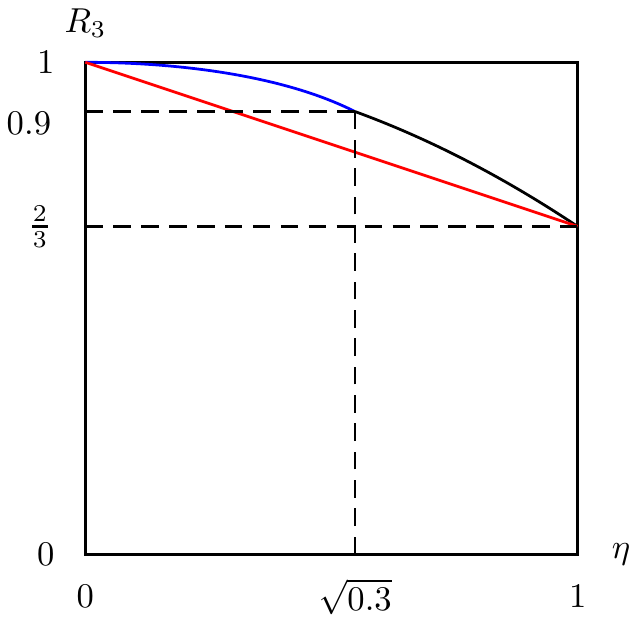}\hskip-2cm\raisebox{4cm}{a}\hfill\raisebox{0.3cm}{\includegraphics[scale=0.85]{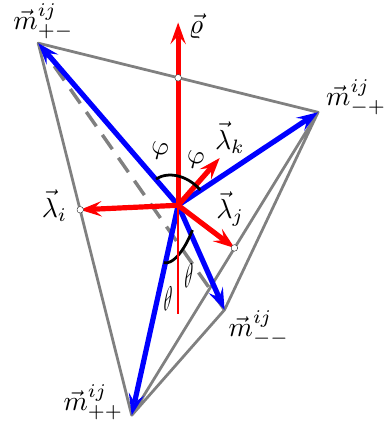}}\hskip-1cm\raisebox{4cm}{b}\hskip1cm\hfill
\caption{(Color online) a) The upper bound of the average anti-correlation $R_3$ in the case of a non-contextual realistic mode, shown as red line, and its violation by a qubit, shown as blue and black curve. b) The optimal pairwise joint measurement $M_{\mu\nu}^{ij}$ of trine spin observables leading to the largest violation with $\varphi=\arctan \sqrt 3\eta_c/\sqrt{H_c}\approx45.74^\circ$ and $\theta=\arctan\eta_c/\sqrt{H_c}\approx30.65^\circ$ with $\eta_c\approx 0.456619$.}
\end{figure}

{\it Theorem 3. For three pairwise jointly measurable unbiased observables $\{O_\pm(\vec\lambda_i)\}$ with the same unsharpness $|\vec\lambda_i|=\eta$ for $i=1,2,3$, it holds
\begin{equation}\label{t3}
R_3\le\left\{\begin{array}{ll}\frac12+\frac{\eta^2}4+\frac12\sqrt{1-2\eta^2+\frac{\eta^4}4},&\eta\le\sqrt{0.3},\\
1-\frac13\eta^2,&\eta\ge\sqrt{0.3}.\end{array}\right.
\end{equation}
The upper bound is attained by trine spin observables, i.e., $\vec\lambda_i\cdot\vec\lambda_j=-1/2$, in the case of $\eta\le\sqrt{0.3}$ and three parallel observables, i.e., $\vec\lambda_1=\vec\lambda_2=-\vec\lambda_3$, in the case of $\eta\ge\sqrt{0.3}$. The optimal state is a pure state whose Bloch vector $\vec\varrho=\tr(\varrho\vec\sigma)$ is orthogonal to $\vec\lambda_{1,2,3}$. The optimal pairwise joint measurements are  composed of four rank-1 effects
\begin{equation}\label{m2}
M_{\mu\nu}^{ij}=\frac{(1+\mu\nu\vec\lambda_i\cdot\vec\lambda_j)(I+\vec m_{\mu\nu}^{ij}\cdot\vec\sigma)}4
\end{equation}
where $\vec m_{\mu\nu}^{ij}\propto \mu\vec\lambda_i+\nu\vec\lambda_j-\mu\nu\vec \varrho\sqrt {H_{ij}}$ are unit vectors  for $i<j$ and $i,j=1,2,3$. }

Proof is given in Supplemental Material \cite{sm}. 
As shown in Fig.1a  LSW's inequality is violated as long as $\eta\not= 0,1$. The maximal violation $\delta=R_3-1+\eta/3$ to LSW's inequality is found numerically to be attained at $\eta_c\approx0.456619$ with $R_3\approx0.937439$
by trine spin observables.  To attain this optimal value, three observables are triplewise jointly measurable with incompatible pairwise joint measurements. The optimal pairwise joint measurements in this case are illustrated in Fig.2b.
In order to further investigate the relevance of triplewise joint measurability to the violation to LSW's inequality, we shall derive the condition for the triplewise joint measurability of three unbiased observables.

The joint measurement of three unbiased orthogonal observables was first considered by Busch \cite{busch2} with a sufficient condition that is proved by Barnnet \cite{ba} to be also necessary. Liang, Spekkens, and Wiseman \cite{lsw} provided  the necessary and sufficient condition for trine spin observables. Pal and Ghosh \cite{pg} proved a necessary condition for the triplewise joint measurability of three general unsharp observables in terms of the Fermat-Toricelli (FT) vector. By definition, a FT vector of a set of three or more vectors $\{\vec v_a\}$ in Euclidean space  is the  vector $\vec v$  that minimizes the total distances $\sum_a|\vec v_a-\vec v|$. The FT vector always exists and is unique \cite{ft}. Pal and Ghosh's necessary condition  \cite{pg} can be proved (Supplemental Material \cite{sm}) to be sufficient for three unbiased observables:

{\it Theorem 4. Three unbiased observables $\{O_\pm(\vec\lambda_k)\}_{k=1}^3$   are triplewise jointly measurable if and only if
\begin{equation}\label{jmc}
\sum_{a=0}^3\left|\vec\Lambda_a-\vec\Lambda_{\rm FT}\right|\le 4
\end{equation}
where $\vec\Lambda_{\rm FT}$ denotes the FT vector of four vectors
\begin{equation}
\vec\Lambda_0=\vec\lambda_1+\vec\lambda_2+\vec\lambda_3,\quad \vec\Lambda_k=2\vec\lambda_k-\vec\Lambda_0\ (k=1,2,3).
\end{equation}}%

The FT vector of  four general vectors does not have an analytical expression. In some special cases such as co-planar vectors and one vector being orthogonal to other two vectors the FT vector can be found explicitly.

\begin{figure}
\includegraphics[scale=0.75]{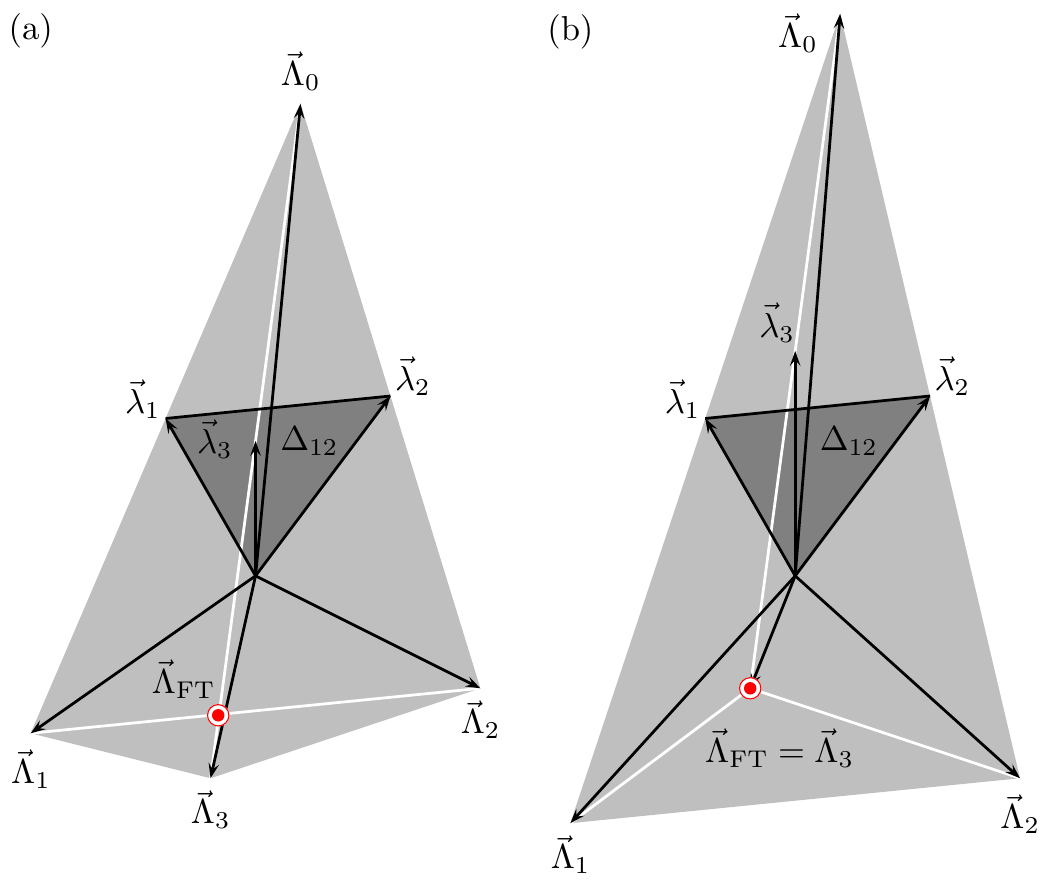}
\caption{(Color online) The FT vector of two different cases of three coplanar observables: a) one observable, e.g., $\vec\lambda_3$, lies in the triangle $\Delta_{12}$ formed by the other two observables, e.g., $\vec\lambda_{1,2}$, and zero vector and b) none of the three observables lies in the triangle formed by the other two observables.}
\end{figure}
{\it Example 1: coplanar observables.} Without loss of generality we can suppose tha three coplanar directions $\{\vec\lambda_i\}$ lie in the same half plane and $\vec\lambda_3$ lies between directions $\vec\lambda_1$ and $\vec\lambda_2$ as shown in Fig.1(a) and Fig.1(b), in which  dark gray shaded region $\Delta_{12}$ denotes the triangle formed by vectors $\vec\lambda_{1,2}$ and the zero vector.    
If $\vec\lambda_3\in \Delta_{12}$  then four vectors $\{\vec\Lambda_a\}_{a=0}^3$
 form a convex quadrilateral as shown in Fig.1(a) and the FT vector can be easily found to be the intersection of two diagonals, shown as red dots in Fig.1. The condition of triplewise joint measurability Eq.(\ref{jmc}) turns out to be exactly the condition of the joint measurability of two observables $\{O_\pm(\vec\lambda_{1,2})\}$. That means the observable that is a convex combination of two jointly measurable observables can also be measured jointly for free. If $\vec\lambda_3\not\in \Delta_{12}$ then vector $\vec\Lambda_3$ falls in the triangle formed by three other vectors and coincides with the FT vector. As a result the condition Eq.(\ref{jmc}) becomes 
\begin{equation}
|\vec\lambda_1+\vec\lambda_2|+|\vec\lambda_1-\vec\lambda_3|+|\vec\lambda_2-\vec\lambda_3|\le 2.
\end{equation}
In particular if all three coplanar vectors $\vec\lambda_k$ have the same length $\eta=|\vec\lambda_k|$ then we have $\vec\lambda_3\not\in \Delta_{12}$. Denoting by $\phi_k$ the angles spanned by $\vec\lambda_3$ and $\vec\lambda_k$ for $k=1,2$,  the triplewise joint measurability condition Eq.(\ref{jmc}) becomes
\begin{equation}
\eta\le\left(\cos\frac{\phi_1+\phi_2}2+\sin\frac{\phi_1}2+\sin\frac{\phi_2}2\right)^{-1}.
\end{equation}
In the case of trine spin observables where $\phi_1=\phi_2=\pi/3$ we reproduce the known condition $\eta\le \frac23$ \cite{lsw}. 

{\it Example 2: $\vec\lambda_3\perp\vec\lambda_{1,2}$.} In this case the the FT vector of four vectors $\{\vec\Lambda_a\}$ can be found explicitly 
$\vec\Lambda_{\rm FT}\propto \vec\lambda_3$ (see Supplemental Material \cite{sm}).
The triplewise joint measurability condition Eq.(\ref{jmc}) now becomes
\begin{equation}\label{lo}
|\vec\lambda_1+\vec\lambda_2|+|\vec\lambda_1-\vec\lambda_2|\le 2\sqrt{1-|\vec\lambda_3|^2}.
\end{equation}
In the case of $\vec\lambda_{1,2}$ also being orthogonal we reproduce the known condition $\sum_i|\vec\lambda_i|^2\le1$ \cite{ba} for the joint measurement of three unbiased orthogonal observables.


Interestingly, there are three observables that are not triplewise jointly measurable but cannot violate LSW's inequality no matter how each two observables are jointly measured. For example, we consider three co-planar observables $\{O_\mu(\vec\lambda_k)\}$ with identical sharpness $\eta =1/\sqrt2 $ and suppose that the angles spanned by $\vec\lambda_3$ and $\vec\lambda_{1,2}$ are $\phi_1=\phi_2=3\pi/4$, respectively. These three observables are obviously pairwise jointly measurable but not triplewise jointly measurable according to condition Eq.(\ref{jmc}). In this case the maximal average anti-correlation reads
$R_3=(3+\sqrt2)/6<1-\sqrt2/6=1-\eta/3$.

In conclusion, by modeling unsharp measurement and joint measurements in  realistic models, we have established LSW's inequality based rigorously and solely on the assumption of non-contextuality. Thus LSW's inequality can be regarded as a genuine KS inequality involving three observables and can be put to experimental tests. The introduction of unsharp and joint measurements in a realistic model allows the room for some kind of measurement contextuality even classically. However this kind of measurement contextuality is not enough to explain all the quantum mechanical predictions on two-level systems as LSW's inequality can be violated, showing that even for a two-level system the attribution of predetermined values to observables  is at odds with quantum mechanics. The maximal violation to LSW's inequality is found to be attained by three pairwise jointly measurable observables that are pairwise measured in an incompatible way. Also we derive the necessary and sufficient condition for the joint measurement of three unbiased observables and show that there are three observables that are not triplewise joint measurable cannot give rise to a violation of LSW's inequality.

This work is funded by the Singapore Ministry of Education (partly through the Academic Research Fund Tier 3 MOE2012-T3-1-009).

\mbox{}\vskip 3cm\mbox{}

\newpage
\setcounter{equation}{0}
\renewcommand{\theequation}{s.\arabic{equation}}
\section*{Supplemental Material}
To recapitulate, in a non-contextual realistic model, all observables possess predetermined values that can be revealed by either a sharp measurement or an unsharp measurement. While a sharp measurement always gives the correct response, unsharp measurement may give wrong response  to the predetermined values. A general unsharp measurement is characterized by a response function satisfying three conditions 
 \begin{equation}
\mbox{RF1. }\tilde\chi_\mu[A]\ge0,\quad \mbox{RF2. }\sum_\mu\tilde\chi_\mu[A]=1
\end{equation}
such that {\it the probability of obtaining outcome $k$ is given by the average of the response function}, i.e.,
\begin{equation}
\mbox{RF3. }P(\mu|\tilde A)=\langle \tilde\chi_\mu[A]\rangle.
\end{equation}

A {\it joint measurement}  outputs a joint probability distribution of measurement results of $A_1$ and $A_2$ in any distribution of the hidden variables $\varrho_\lambda$. Thus it has  a response function $\tilde\chi_{\mu\nu}[A_{12}]$ with two response functions of the given measurements of $A_1$ and $A_2$ as marginals
\begin{equation}
\tilde\chi_{\mu}[A_1]=\sum_\nu\tilde\chi_{\mu\nu}[A_{12}],\quad \tilde\chi_{\nu}[A_2]=\sum_\mu\tilde\chi_{\mu\nu}[A_{12}].
\end{equation}
Three pairwise joint measurements  $\tilde\chi_{\mu\nu}[A_{jk}]$  of three observables $A_k$ with $j< k$ and $j,k=1,2,3$ are   {\it compatible} if there exists a joint measurement $\tilde\chi_{\mu\nu\tau}[A_{123}]$ having those three response functions of pairwise joint measurements as marginals, e.g., 
\begin{equation}
\tilde\chi_{\mu\nu}[A_{12}]=\sum_\tau\tilde\chi_{\mu\nu\tau}[A_{123}].
\end{equation} 

Specifically, a general measurement of a binary observable $A_k$ taking values $\pm$ has the following response function
 \begin{equation}
\tilde\chi_\mu[A_k]=\frac{1+\mu\eta_k(\lambda) A_k}2:=\frac{1+\mu\tilde A_k}2.
\end{equation} 
The most general joint measurement of two observables $A_j$ and $A_k$  has the following response function
\begin{equation}\label{srf2}
\tilde\chi_{\mu\nu}[A_{jk}]=\frac{1+\mu\eta_j(\lambda)A_j+\nu\eta_k(\lambda)A_k+\mu\nu C_{jk}}4
\end{equation}
where $C_{jk}$ is an arbitrary observable whose predetermined values satisfy
$
1\pm C_{jk}\ge| \eta_j(\lambda) A_j\pm\eta_k(\lambda)A_k|
$. 
The {\it (global) sharpness} $\eta$ of an unsharp measurement can be understood in a theory independent fashion as 
\begin{equation}
\eta=\min\max_{A=\pm}\left|P(+|\tilde A)-P(-|\tilde A)\right|
\end{equation}
with minimization taken over all possible states in which observable $A$ has definite values. In a realistic model $\eta=\min\langle |\eta(\lambda)|\rangle$ over all distributions $\varrho_\lambda$.

{\it Proof of Theorem 1. ---} 
 Consider three observables $A_k$ each of which is measured unsharply with some sharpness $\eta_k(\lambda)$  and  response functions 
$\tilde\chi_\mu[A]$ for $k=1,2,3$. Let $\tilde\chi_{\mu\nu}[A_{jk}]$ be the response function of three pairwise joint measurements for $j<k$. If these three pairwise joint measurements are compatible then there exists
a triplewise joint measurement with  response function 
\begin{eqnarray}\label{cjm3}
8\tilde\chi_{\mu\nu\tau}[A_{123}]=1+\mu\tilde A_1+\nu\tilde A_2+\tau\tilde A_3\quad\quad\quad\nonumber\\
+\mu\nu C_{12}+\mu\tau C_{13}+\nu\tau C_{23}+\mu\nu\tau C,
\end{eqnarray}
where $C$ is an arbitrary observable whose predetermined values must ensure $\tilde\chi_{\mu\nu\tau}[A_{123}]\ge0$, which are equivalent to $1+\tau C\ge \Gamma_{\mu\nu}^\tau$, for all $\mu,\nu,\tau=\pm$ where
\begin{eqnarray}
 \Gamma_{\mu,\nu}^\tau= \mu(\tilde A_1+\tau C_{23})+\nu(\tilde A_2 +\tau C_{13})\quad\nonumber\\-\mu\nu(C_{12}+\tau \tilde A_3).
\end{eqnarray}
As a result we obtain 
\begin{equation}\label{gma}
2\ge \Gamma^+_{\mu,\nu}+\Gamma^-_{\mu\mu^\prime,\nu\nu^\prime},\quad (\mu,\nu,\mu^\prime,\nu^\prime=\pm).
\end{equation} 
In the case of $\mu^\prime=\nu^\prime=-$, we obtain $$1+\mu C_{12}+\nu C_{23}+\mu\nu C_{13}\ge 0,\quad (\mu,\nu=\pm)$$
from which it follows Eq.(\ref{t1}).
On the other hand, by noticing that Eq.(\ref{gma}) is ensured  by $\tilde\chi_{\mu\nu}[A_{jk}]\ge0$ for $j<k$ in the case of $(\mu^\prime,\nu^\prime)\not=(-,-)$  and by the condition Eq.(\ref{t1}) in the case of $(\mu^\prime,\nu^\prime)=(-,-)$, the choice
$$C=\max_{\mu,\nu}\Gamma_{\mu,\nu}^+-1$$
makes Eq.(\ref{cjm3}) a response function of a triplewise joint measurement with three given pairwise joint measurements as marginals, i.e., they are compatible.\hfill$\square$

{\it Proof of Theorem 2. ---} By substituting the response functions  Eq.(\ref{rf2}) of the most general pairwise measurements into $R_3$ we obtain
\begin{eqnarray}
R_3&=&\frac13\sum_{j<k}\big({P(+,-|A_{jk})+P(-,+|A_{jk})}\big)\nonumber\\
&=&\frac13\sum_{j<k}\langle{\tilde\chi_{+-}[A_{jk}]+\tilde\chi_{-+}[A_{jk}]}\rangle\label{r31}\\
&=&\frac13\sum_{j<k}\frac{1-\langle C_{jk}\rangle}2=1-\frac13\sum_{j<k}\frac{1+\langle C_{jk}\rangle}2\nonumber\\
&\le&1-\frac16\sum_{j<k}{\langle|\eta_j(\lambda) A_j+\eta_k(\lambda)A_k|\rangle}\label{r32}\\
&\le&1-\frac13\langle \max_k|\eta_k(\lambda)|\rangle\label{r33}\\&\le&
1-\frac13\max_k\langle |\eta_k(\lambda)|\rangle\le 1-\frac13\max_k\eta_k.\label{r34}
\end{eqnarray}
Here Eq.(\ref{r31}) is due to the defining property RF3 of response function and the first inequality Eq.(\ref{r32}) is due to  $\chi_{\mu}[A_{jk}]\ge0$ while the second inequality Eq.(\ref{r33}) is due to the triangle inequality  $$\sum_{j<k}|\eta_j(\lambda) A_j+\eta_k(\lambda)A_k|\ge2|\eta_i(\lambda)|,$$  considering $|A_i|=1$, for any $i=1,2,3$. The last inequality Eq.(\ref{r34}) is due to the definition of the global sharpness $\eta_k=\min\langle|\eta_k(\lambda)|\rangle\le \langle|\eta_k(\lambda)|\rangle$.
\hfill$\square$

{\it Proof of Theorem 3. ---} For given two unbiased observables $\{O_\mu(\vec\lambda_{i,j})\}$ that are jointly measurable, i.e., $H_{ij}\ge0$ the most general joint measurement is given by
\begin{equation}\label{j2}
M_{\mu\nu}^{ij}=\frac{I+\mu\nu Z_{ij}+(\mu\vec\lambda_i+\nu\vec\lambda_j-\mu\nu\vec z_{ij})\cdot\vec\sigma}4
\end{equation}
with real number $Z_{ij}$ and vector $\vec z_{ij}$ making $M_{\mu\nu}^{ij}\ge0$ for all $\mu,\nu=\pm$. This positivity requirement is equivalent to
\begin{equation}
|\vec z_{ij}|^2\le{(1+\mu Z_{ij})^2-|\vec\lambda_i+\mu\vec\lambda_j|^2}:=L_\mu(Z_{ij})
\end{equation}
for $\mu=\pm1$. Obviously the condition is necessary for $M_{\mu\nu}^{ij}\ge0$. To show its sufficiency we note that for each allowed value of $Z_{ij}$ determined by $L_\mu(Z_{ij})\ge0$, we choose $\vec z_{ij}$ to be a vector orthogonal to both $\vec\lambda_{i,j}$ with a length $\min_\mu L_\mu(Z_{ij})^{1/2}$, which define a joint measurement via Eq.(\ref{j2}). 
 In a given state $\varrho$ with a Bloch vector $\vec\varrho=\tr\vec\sigma\varrho$ the anti-correlation for a given joint measurement Eq.(\ref{j2}) reads 
\begin{eqnarray}
R_{ij}&=&\tr\varrho(M_{+-}^{ij}+M_{-+}^{ij})\nonumber\\&=&\frac{1-Z_{ij}+\vec\varrho\cdot\vec z_{ij}}2\le \frac{1-Z_{ij}+|\vec z_{ij}|}2\nonumber\\
&\le &\frac{1-Z_{ij}+\min_{\mu}\sqrt{L_\mu(Z_{ij})}}2\nonumber\\
&\le &\frac{1-\vec\lambda_i\cdot\vec\lambda_j+\sqrt{H_{ij}}}2
\end{eqnarray}
The last inequality is due to the fact that $\min_\mu L_{\mu}(Z_{ij})\le H_{ij}$ in the case of $Z_{ij}\ge \vec\lambda_i\cdot\vec\lambda_j$ and $-Z_{ij}+\sqrt{L_{+}(Z_{ij})} $ is an increasing function of $Z_{ij}$ in the case of $Z_{ij}\le\vec\lambda_i\cdot\vec\lambda_j$. 

In the case of identical sharpness $|\vec\lambda_k|=\eta$ we denote $\vec\lambda_i\cdot\vec\lambda_j=\eta^2x_k$ for $(i,j,k)$ being three cyclic permutation of $(1,2,3)$. As a result the average anti-correlation has the following upper bound 
\begin{eqnarray}\label{u}
R_3\le \frac{r(x_1)+r(x_2)+r(x_3)}6
\end{eqnarray}
where
$
r(x)={1-\eta^2x+\sqrt{1-2\eta^2+\eta^4x^2}}.
$
At least two out of three $x_k$ should be negative to achieve the largest upper bound. Without loss of generosity we suppose $x_3\le 0$.
Taking into account the positive semi-definiteness of  the Gram matrix $[[\vec\lambda_i\cdot\vec\lambda_j]]\ge0 $ for three vectors $\vec\lambda_k$, it holds $x_3\ge x_1x_2- \bar x_1\bar x_2$
where $\bar x_k=\sqrt{1-x_k^2}$. Since the function $r(x)$ is a decreasing function of $x$ in the case of $x\le 0$, the upper bound achieves the largest value when $x_3=x_1x_2- \bar x_1\bar x_2$ and in this case we denote by $U(x_1,x_2)$ the r.h.s. of Eq.(\ref{u}). This upper bound is actually attained by coplanar observables as shown in \cite{kg}.

The maximal value of $U(x_1,x_2)$ is either achieved at the boundary $x_{1,2}=\pm1$ or at the critical points  determined by $\partial_k U(x_1,x_2)=0$ for $k=1,2$, or equivalently
\begin{eqnarray}
\bar x_k r^\prime(x_k)+({x_1\bar x_2+x_2\bar x_1})r^\prime(x_3)=0.
\end{eqnarray}
It turns out that for any given constant $r$ the equation $r=\bar x r^\prime (x):=\tilde r(x)$  has at most two solutions. This conclusion follows from the fact that the equation $\tilde r^\prime (x)=0$,  which is equivalent to
\begin{equation}
x(\eta^2 x-\sqrt{1-2\eta^2+\eta^4x^2})=\frac{{\eta^2(1-2\eta^2)}(1-x^2)}{{1-2\eta^2+\eta^4x^2}}
\end{equation}
for $x\not=\pm1$, has at most one solution. In fact,
in the case of $1-2\eta^2>0$, the equation has only nonpositive solutions and l.h.s. is an increasing function of $-x$ while the r.h.s. is a decreasing function of $-x$. Thus there is at most one solution in this case and at most two solutions to the equation $\tilde r(x)=r$. In the case of $1-2\eta^2<0$ the l.h.s.$\ge0$ while the r.h.s.$<0$ so that  $\tilde r^\prime (x)=0$ has no solution which means $\tilde r(x)=r$ has at most one solution for any $r$. 

By noting that $\bar x_3=|{x_1\bar x_2+x_2\bar x_1}|$ and the fact that the critical points of $U(x_1,x_2)$ are determined by $\tilde r(x_1)=\tilde r(x_2)=\alpha \tilde r(x_3)$ with $\alpha=-sgn (x_1\bar x_2+x_2\bar x_1)$, we can conclude that at least two out three $x_{1,2,3}$ 
must be equal and negative. Therefore we have the upper bound
$$R_3\le\max_{-1\le x\le 0}\frac{g(x)}6,\quad g(x)=2r(x)+r(2x^2-1).$$
Function $g(x)$ is well defined in the interval a)  $[-1,0]$ if $\eta\le 1/\sqrt2$; b)  $I_1\cup I_2$ if $\sqrt3-1\ge \eta> 1/\sqrt2$; c) $I_1$, if $\eta>\sqrt3-1$ where
$$I_1:=\left[-1, -\sqrt{\frac{1+\beta}2}\right],\quad I_2=\left[-\sqrt{\frac{1-\beta}2}, -\beta\right]$$ with $\beta={\sqrt{2\eta^2-1}}/{\eta^2}$. In case c) function $g(x)$ 
is decreasing so that its maximum is attained at $x=-1$. In case b) function $g(x)$ is monotonously increasing in the interval $I_1$ and concave in $I_2$ with $x=-1/2$ as the unique critical point. Thus its maximum is achieved at either $x=-1$ or $x=-1/2$. In case a), if $\eta<1/2$ function $g(x)$ has  a unique critical point $x=-1/2$. If $1/2\le \eta\le 1/\sqrt 2$ function $g(x)$ has two critical points with $x=-1/2$ being local maximum and the other being local minimum.  As a result the maximal value of $g(x)$ is attained either by the trine spin observables, i.e., $x=-1/2$ or collinear observables $x=-1$. All these properties can be checked  for the function $g(x)$ of single variable numerically and ultimate upper bound is given by Eq.(\ref{t3}) which is depicted in Fig.1a. On the other hand it is straightforward to check that the optimal pairwise joint measurements given in Eq.(\ref{m2}) and state specified in Theorem 3 actually attain the upper bound.\hfill$\square$

{\it Proof of Theorem 4. ---}The necessary part has already been proved by Pal and Ghosh \cite{pg} in the case of three general unsharp observables. 
Here we include its proof for unbiased observables  for the sake of completeness. The most general form of triplewise joint measurement, if exists, takes the following form
\begin{eqnarray}\label{m3}
8M_{\vec\mu}=I+\sum_{i>j}{\mu_i\mu_j} (Z_{ij}+\vec z_{ij}\cdot\vec\sigma)\hskip 1cm\nonumber\\
+\sum_{i=1}^3{\mu_i}\vec\lambda_i\cdot\vec\sigma-{\mu_1\mu_2\mu_3}\vec z\cdot\vec\sigma
\end{eqnarray}
with real constants $Z_ij$ and vectors $\vec z_ij$ and $\vec z$ making $M\ge0$, which is equivalent to
\begin{eqnarray}\label{mp}
1+\sum_{i>j}\mu_i\mu_jZ_{ij}\hskip 4cm\nonumber\\\ge
\Big|\sum_i\mu_i\vec\lambda_i+\sum_{i>j}\mu_i\mu_j\vec z_{ij}-\mu_1\mu_2\mu_3\vec z\Big|,
\end{eqnarray}
By summing over all $\mu_k=\pm1$ and separating two different cases $\mu_1\mu_2\mu_3=\pm1$ we obtain
\begin{eqnarray}
8&\ge&\sum_{\mu}\Big|\sum_i\mu_i\vec\lambda_i-\sum_{i>j}\mu_i\mu_j\vec z_{ij}-\mu_1\mu_2\mu_3\vec z\Big|\cr
&=&\sum_{\mu_1\mu_2\mu_3=1}\Big|\sum_i\mu_i\vec\lambda_i-\sum_{i>j}\mu_i\mu_j\vec z_{ij}-\vec z\Big|+\cr
&&\sum_{\mu_1\mu_2\mu_3=-1}\Big|-\sum_i\mu_i\vec\lambda_i-\sum_{i>j}\mu_i\mu_j\vec z_{ij}+\vec z\Big|\cr
&\ge&2\sum_{\mu_1\mu_2\mu_3=1}\Big|\sum_i\mu_i\vec\lambda_i-\vec z\Big|\cr
&\ge&2\sum_{\mu_1\mu_2\mu_3=1}\Big|\sum_i\mu_i\vec\lambda_i-\vec \Lambda_{\rm FT}\Big|
\end{eqnarray}
with the last inequality due to the definition of the FT vector $\vec\Lambda_{\rm FT}$ of four vectors $\{\sum_i\mu_i\vec\lambda_i\mid \mu_1\mu_2\mu_3=1\}$.

To prove its sufficiency we consider eight operators $M_{\vec\mu}$ as given in Eq.(\ref{m3}) with $\vec z=\vec\Lambda_{\rm FT}$, $\vec z_{ij}=0$, and
 \begin{equation*}
Z_{ij}=1-\frac{|\vec\Lambda_i-\vec\Lambda_{\rm FT}|+|\vec\Lambda_j-\vec\Lambda_{\rm FT}|}2
\end{equation*} 
with $i>j$ and $i,j=1,2,3$.
It is obvious that three given unbiased observables arise as marginals of $\{M_{\vec\mu}\}$, e.g., $O_{\mu_1}(\vec\lambda_1)=\sum_{\mu_2\mu_3}M_{\vec\mu}$ and the conditions Eq.(\ref{mp}), which are equivalent to $M_{\vec\mu}\ge 0$, are ensured by Eq.(\ref{jmc}) for  $\vec\mu=(\pm\pm\pm)$ and become equalities otherwise. \hfill$\square$

{\it FT vector in the case of $\vec\lambda_3\perp\vec\lambda_{1,2}$. --- } With the help of Lindloef and Sturm  condition, which in this case reads 
\begin{equation*}\label{ft2}
\sum_{a=0}^4\frac{\vec\Lambda_a-\vec\Lambda_{\rm FT}}{|\vec \Lambda_a-\vec\Lambda_{\rm FT}|}=0
\end{equation*}
we can find that 
\begin{equation*}
\vec\Lambda_{\rm FT}=\frac{|\vec\lambda_1+\vec\lambda_2|-|\vec\lambda_1-\vec\lambda_2|}{|\vec\lambda_1+\vec\lambda_2|+|\vec\lambda_1-\vec\lambda_2|}\vec\lambda_3.
\end{equation*}
\newpage
\end{document}